\documentclass[5p]{elsarticle}

\usepackage{import}
\usepackage{amsmath}
\usepackage{amsfonts}
\usepackage{amssymb}
\usepackage{ragged2e}
\usepackage{multicol}
\usepackage{lscape}
\usepackage{xcolor}
\usepackage{soul}
\usepackage[utf8]{inputenc}
\usepackage[T1]{fontenc}
\usepackage{lmodern}
\usepackage{siunitx}
\usepackage{textcomp}
\usepackage{upgreek}
\usepackage{stfloats}
\usepackage{blindtext}
\usepackage{physics}
\numberwithin{equation}{section}
\usepackage{times}
\usepackage{setspace}
\usepackage{import}
\usepackage{enumitem}
\usepackage{etoolbox} 
\usepackage{lipsum} 
\usepackage{mathtools}
\usepackage{float}

\usepackage{orcidlink}

\begin{document}
\begin{frontmatter}

\title{Topology optimization of cathode gas channel layout in advanced proton exchange membrane fuel cells}

\author[1]{Zahra Kazemi\corref{cor1}\orcidlink{https://orcid.org/0000-0002-4325-0101}}

\ead{zahra.kazemi@mail.utoronto.ca}

\affiliation[1]{organization={Advanced Research Laboratory for Multifunctional Lightweight Structures (ARL-MLS), Department of Mechanical and Industrial Engineering, University of Toronto}, 
    city={Toronto},
    postcode={M5S 3G8}, 
    province={ON,},
    country={Canada}}

\author[1]{Kamran Behdinan\corref{cor1}\orcidlink{https://orcid.org/0000-0002-1873-9837}}
\ead{kamran.behdinan@utoronto.ca}

\cortext[cor1]{Corresponding author}

\onecolumn




\newpage

\begin{abstract}
The proton exchange membrane fuel cell (PEMFC) output relies on the transport behavior within the cathode gas channels. Current designs remain inadequate as they often rely on heuristic modifications of existing layouts or designer intuition with suboptimal performance. In this study, topology optimization is proposed to redesign the PEMFC cathode gas channel layout without a priori assumptions. The optimization aims to maximize the reactant concentration and minimize power dissipation along the flow path. The problem is solved within a three-dimensional half-cell model. For computational tractability, a reduced-order, depth-averaged two-dimensional model is also implemented. The optimized topology yields an enhanced current density with lower energy dissipation over the conventional benchmarks. At an inlet velocity of 0.15 m/s, the pressure drop is reduced by 46.7\% compared to the serpentine layout, though is 28.2\% higher than that of the parallel case. Within the optimized channels, oxygen flows at higher local velocities, which allows a more homogeneous reactant delivery across the domain. Relative to the serpentine layout, the improvement in mean current density reaches 20.9\% with 4.9\% lower standard deviation. Placing more emphasis on dissipation minimization during optimization produces more intricate, tortuous channel topologies. Such design flexibility enables the discovery of unconventional yet efficient layouts.
\end{abstract}

\begin{keyword}
Fuel cell \sep Flow channel route \sep Three-dimensional topology optimization \sep Power dissipation reduction \sep Homogenized reactant distribution 
\end{keyword}

\end{frontmatter}

\onecolumn

\begin{multicols}{2}
  \begin{tabbing}
    \hspace{2cm}\= \hspace{8cm}\= \kill
    \textbf{\small Nomenclature} \\[0.5em]
    $L$ \> membrane length \\
    $W$ \> inlet width \\
    $H_{gdl}$ \> depth of GDL \\
    $H_{cl}$ \> catalyst layer thickness \\
    $H_{mem}$ \> membrane depth \\
    $H_{bp}$ \> channel height \\
    $\Omega$ \> design domain \\
    $\Gamma_{in}$ \> inlet boundary \\
    $\Gamma_{out}$ \> outlet boundary \\
    $\Gamma_{sym}$ \> symmetry boundary \\
    $\rho$ \> gas density \\
    $u_i$ \> velocity in the $i$-th direction \\
    $x_i$ \> spatial coordinate \\
    $\varepsilon$ \> medium porosity \\
    $p$ \> pressure \\
    $F_{b_i}$ \> flow friction force in the $i$-th direction \\
    $\alpha_b$ \> design-dependent Brinkman constant \\
    $\kappa$ \> permeability of the medium \\
    $\mu$ \> gas dynamic viscosity \\
    $\omega_k$ \> mass fraction of the $k$-th species \\
    $j_{k_i}$ \> diffusion mass flux of the $k$-th species \\
    $n_s$ \> number of chemical species \\
    $f_D$ \> diffusivity factor \\
    $D_{kj}$ \> Maxwell-Stefan diffusivity tensor \\
    $D^m_{k}$ \> diffusion coefficient \\
    $D^{ref}_{kj}$ \> reference diffusivity tensor \\
    $M^m$ \> mean molar mass \\
    $M_k$ \> molar mass \\
    $R_k$ \> mass generation rate \\
    $\nu_k$ \> stoichiometric coefficient of the reaction \\
    $n_e$ \> number of participating electrons \\
    $a_v$ \> active specific area \\
    $F$ \> Faraday constant \\
    $i_{loc}$ \> local current density \\
    $\eta_c$ \> cathode overpotential \\
    $\alpha_c$ \> cathode transfer coefficient \\
    $\alpha_a$ \> anodic transfer coefficient \\
    $T$ \> cell temperature \\
    $T_{ref}$ \> reference temperature \\
    $p_{ref}$ \> reference pressure \\
    $R_g$ \> universal gas constant \\
    $i_0$ \> volumetric exchange current density \\
    $V_{eq}$ \> equilibrium voltage \\
    $V_{cell}$ \> prescribed cell voltage \\
    $R$ \> charge transport resistance \\
    $u_{in}$ \> inlet velocity \\
    $n$ \> outgoing unit normal vector \\
    $\gamma$ \> material indicator (design variable) \\
    $\kappa_{\text{c}}$ \> effective permeability of the channel \\
    $\kappa_{\text{r}}$ \> effective permeability of the rib \\
    $q_{\alpha}$ \> convexity factor in interpolation scheme \\
    $q_f$ \> SIMP penalization exponent \\
    $f_{D_r}$ \> rib diffusivity correction factor \\
    $f_{D_c}$ \> channel diffusivity correction factor \\
    $C(x)$ \> reactant concentration field \\
    $\overline{C}$ \> mean reactant concentration \\
    $\sigma_{c}$ \> standard deviation of concentration field \\
    $\mathbf{\gamma}$ \> vector of design variables \\
    $N_{dof}$ \> number of design variables (degrees of freedom) \\
    $V^{\star}$ \> prescribed admissible volume \\
    $\overline{C}^{\mathrm{init}}$ \> initial mean concentration \\
    $\sigma_c^{\mathrm{init}}$ \> initial std. deviation of concentration field \\
    $\omega$ \> weighting factor in the multi-objective formulation \\
    $J_f$ \> power dissipation \\
    $\gamma_f$ \> filtered design variable \\
    $r_{min}$ \> filter radius \\
    $\gamma_h$ \> projected design variable \\
    $\theta_p$ \> projection point \\
    $\beta$ \> projection slope \\
    $\Delta h$ \> characteristic mesh size \\
    $N_{ele}$ \> number of mesh elements \\
    $V$ \> volume of the meshed computational domain \\
    $d$ \> spatial dimension \\
    $C_{i}$ \> constraint penalty coefficient \\
    $t_{opt}$ \> optimality convergence tolerance \\
    $N_{ev}$ \> maximum number of model evaluations \\
    $N_{in}$ \> maximum number of inner iterations \\[0.5em]

    \textbf{\small Abbreviations} \\[0.5em]
    AFC \> alkaline fuel cell \\
    PAFC \> phosphoric acid fuel cell \\
    PEMFC \> proton exchange membrane fuel cell \\
    MCFC \> molten carbonate fuel cell \\
    SOFC \> solid oxide fuel cell \\
    GDL \> gas diffusion layer \\
    GC \> gas channel \\
    CL \> catalyst layer \\
    AM \> additive manufacturing \\
    SLS \> selective laser sintering \\
    SWFC \> sinusoidal wave channel \\
    OSWFF \> opposite sinusoidal wave field \\
    ORR \> oxygen reduction reaction \\
    SIMP \> solid isotropic material with penalization \\
    HPM \> heaviside projection method \\
    RAMP \> rational approximation of material properties \\
    FEA \> finite element analysis \\
    PDE \> partial differential equation \\
  \end{tabbing}
\end{multicols}

\normalsize

\section{Introduction}

Ecological footprint and rapid, unsustainable depletion of fossil fuels are global concerns that demand worldwide attention. As a renewable energy source, hydrogen gradually reduces reliance on carbon-based energy sources for electricity generation \cite{yue2021hydrogen, wood2025critical}. Fuel cells, which harness the chemical energy of hydrogen to produce electricity, are competitive alternatives for clean and sustainable power production \cite{sorensen2018hydrogen, felseghi2019hydrogen, manoharan2019hydrogen}. The main fuel cell types include solid oxide (SOFC) \cite{gur2016comprehensive}, proton exchange membrane (PEMFC) \cite{tellez2021proton}, alkaline (AFC) \cite{ferriday2021alkaline}, phosphoric acid fuel cell (PAFC) \cite{qasem2024recent}, and molten carbonate (MCFC) \cite{dicks2004molten}. Among these, proton exchange membrane fuel cell (PEMFC) technology has been adopted as a solution for power generation in stationary and vehicular applications \cite{hong2018recent, sando2009research, wee2007applications}, due to its low to zero pollutant emission, compact and space-efficient design, sustained operation at high current densities, low operating temperatures and high practical efficiency. 

PEMFC performance is linked to the multiphase transport characteristics of individual components within the cell stack, especially the gas channels (GCs) placed within the bipolar plate. In particular, the optimization of the gas channel layout remarkably contributes to the proper functioning of the cell. It controls the homogeneity of reactant distribution at active sites to prevent the formation of local hot and cold spots, minimizes pressure drop along the flow path to avoid installation of external blowers for reactant recirculation, and supports the drainage of byproduct liquid water to prevent flooding \cite{kumar2003effect}. This has motivated the majority of prior works in the literature on bipolar plate design enhancement. 



Previous studies can primarily be divided into the optimization of channel geometric parameters and channel arrangements. The latter is considered more influential on plate performance. Conventional channel arrangements, such as serpentine, parallel, interdigitated, and mesh, have been studied experimentally and computationally \cite{manso2012influence, heidary2016experimental, gundlapalli2020performance}. Serpentine and interdigitated designs are favorable for homogeneous reactant transport over the GDL. The serpentine channel design minimizes the risk of water flooding, as the liquid water is pushed out by the reactant stream \cite{lim2016effects}, but it induces a considerable pressure loss. In contrast, the parallel configuration ensures minimal pressure drop with its multiple short parallel routes, but suffers from a poor, uneven reactant distribution \cite{hontanon2000optimisation, lobato2010three}. At high current densities, parallel channels are prone to water clogging that blocks reactant diffusion to the catalyst sites \cite{lim2016effects, gundlapalli2020performance}. Interdigitated designs generally outperform parallel patterns, but not serpentine ones \cite{spernjak2010situ}. Microlattice structures inside the GCs, as used in Toyota fuel cell vehicles, have evolved from mesh layouts and are designed to enhance heat dissipation and electric conductance inside the channels, and improve two-phase flow mechanisms \cite{konno2015development, wang2009porous, park2019achieving, afshari2017investigation}. This enables the use of thinner bipolar plates. The improved convective mass transport was also observed in foam flow field designs \cite{bao2019numerical}. 

Researchers have also taken inspiration from natural flow distribution systems to enhance the multiphase transport properties of bipolar plates \cite{iranzo2020biomimetic, currie2010biomimetic, domachuk2010bio}. Trogadas et al. \cite{trogadas2018lung} introduced a lung-inspired model with multiple branching generations to address reactant homogeneity issues in PEMFCs. Their 3D-printed, four-generation lung-shaped GCs outperformed conventional serpentine designs with up to 30\% increase in maximum power density at current densities above 0.8 $\mathrm{A/cm^2}$. In terms of pressure drop, a 50\% reduction was reported. Similarly, Ozden et al. \cite{ozden2017designing} tested a leaf-inspired cathode flow field paired with a serpentine design on the anode side. This combination yielded a power density of 888 $\mathrm{W/m^2}$, compared to 824 $\mathrm{W/m^2}$ for serpentine channels on both sides. Computational and experimental measurements of selective laser sintered (SLS) leaf-shaped designs showed a notable energy dissipation reduction and comparable output power \cite{guo2011bio}. Wang et al. \cite{wang2021bio} proposed a fishbone-assisted model that enhanced oxygen transport and improved water removal capability from the GDL. Compared to the parallel arrangement, this structure had greater uniformity in reactant and saturation distributions, and thereby enhanced cell operational stability. The geometry was further optimized by adjusting the width and number of sub-branches. Kloess et al. \cite{kloess2009investigation} integrated leaf-inspired features with conventional layouts to design a novel interdigitated pattern. Compared to standard serpentine and interdigitated, such GCs led to a reduced pressure drop (27 Pa vs. 38-41 Pa), homogenized reactant diffusion, and up to 30\% increase in peak power density. Roshandel et al. \cite{roshandel2012simulation} also developed a leaf-venation-inspired layout and reported power density improvements of 56\% and 27\% over parallel and serpentine structures, respectively. Cai et al. \cite{cai2020design} proposed a wave-like structure based on squid fin design, which showcased a high transport efficiency and low flow resistance. Similarly, another study reported that wavy channels outperformed straight ones by a 23.8\% higher power density at 0.4 V. The optimal channel depth and wavelength of 0.45 mm and 2 mm, respectively, were recommended. 

In a comparative study on sinusoidal wave channels (SWFCs), Anyanwu et al. \cite{anyanwu2019comparative} observed enhanced liquid water drainage rate, especially with longer sinusoidal distances \cite{anyanwu2019comparative}. Zhou et al. \cite{zhou2023optimal} proposed opposite sinusoidal wave flow fields (OSWFFs), particularly at higher current densities. They tested a dual-inlet configuration to improve the uniformity of reactant concentration. Multiple inlets and outlets are favorable for uniform distribution and reduced energy dissipation by shortening flow paths \cite{wang2022multi, wang2018numerical}. The dual-inlet OSWFF with a trapezoidal block achieved a 9.3\% increase in power density and over 10\% reduction in liquid water saturation compared to the parallel design. Wave-like channel configurations have also found practical applications in fuel cell electric vehicles \cite{yoshida2015toyota, nonobe2017development, han2014simulation}. 

For rectangular channels, a width-to-depth ratio of 3 has been recommended for optimal performance \cite{kumar2003effect}. Larger ratios enhance reactant distribution and facilitate water removal, but also tend to increase pressure losses and ohmic impedance \cite{wang2017respective, zehtabiyan2017effect}. Furthermore, higher current densities have been reported for shallower channels, with an optimal depth of 1 mm \cite{afshari2017investigation}. Although most prior works have studied rectangular profiles, alternative cross sections, such as hemispherical, triangular, and trapezoidal, have also been explored. Triangular and hemispherical channels were found to be efficient in reactant access to the catalyst layer, but issues related to water behavior were not discussed \cite{kumar2003effect}. Freire et al. \cite{freire2014influence} later confirmed that trapezoidal and triangular channels restrict the buildup of water droplets. Tapered channels with varying cross sections along the streamline direction have also drawn attention. Increasing channel width and decreasing its height along the flow path was shown to enhance overall cell performance \cite{yan2006numerical, liu2006reactant}. Wang et al. \cite{wang2010inverse} optimized tapered serpentine channels with varying heights and reported an $11.9$\% higher output power compared to uniform-height channels. Similarly, Zeng et al. \cite{zeng2017optimization} optimized trapezoidal channels by varying the widths of bases as the optimization variables. The objective function included flow power consumption and cell output power.  At $0.4$ V, an edge ratio of 1.45 produced an $8$\% higher current density than the square-channel design. The optimal design showed a more homogeneous distribution of current density than the basic design. He et al. \cite{he2020novel} added trapezoidal baffles to the channels and examined the relationship between transport performance and inclination angle. They concluded that smaller inclination angles are suitable for high water content cases. Yin et al. \cite{yin2019influence} further showed that the inclusion of baffled conduits improves vertical convective mass transfer. A sloping angle of 45$^\circ$ showcased the highest performance enhancement.

Despite the progress made on bipolar plate design enhancement, current designs remain inadequate, as they often rely on heuristic modifications of existing structures or designer intuition. These methods are typically constrained by a limited set of design variables and a low degree of design freedom that restricts notable design changes. Topology optimization is a promising alternative to conventional heuristic-based design approaches. Initially formulated by Bendsøe and Kikuchi \cite{bendsoe1988generating}, topology optimization is a mathematical approach that identifies the optimal arrangement of material within a design space to maximize a chosen performance metric subject to constraints. It enables a high degree of design freedom and allows for the creation of unconventional and nonintuitive geometries. The geometric freedom offered by AM allows for the fabrication of such designs with intricate features. Topology optimization has been used for convective heat transfer and mass transfer problems \cite{luo2023topology, li2022optimum, li2019optimal, alexandersen2020review, borrvall2003topology}. However, its application to flow field design in PEMFCs remains relatively underexplored. This study aims to explore topology optimization for PEMFC flow field design without a priori assumptions on the layout. In particular, the cathode GC layout is optimized to enhance air transport capacity at minimized pressure drop. The optimization problem is initially solved in three dimensions within a half-cell PEMFC model to optimize the cathode GCs. However, 3D topology optimization is prone to local minima and a high computational cost. To enhance numerical stability and enable a broader design exploration, a reduced-order, depth-averaged two-dimensional model is also implemented inspired from \cite{he2019reduced, behrou2019topology}. A gradient-based optimizer, with gradients derived from the discrete adjoint method, is used for design optimization. The resulting topology-optimized GCs are then analyzed and evaluated against conventional reference designs.

The rest of the article continues with Section 2, which outlines the underlying physics and the governing equations for the PEMFC system. Section 3 introduces the topology optimization problem. Section 4 showcases the numerical results and discusses the proposed designs compared with standard baseline configurations. Lastly, Section 5 concludes with a summary of the outcomes and possible avenues for future research.

\section{Physical Model}
\begin{figure*}[h!]
    \centering
    \includegraphics[width=12cm]{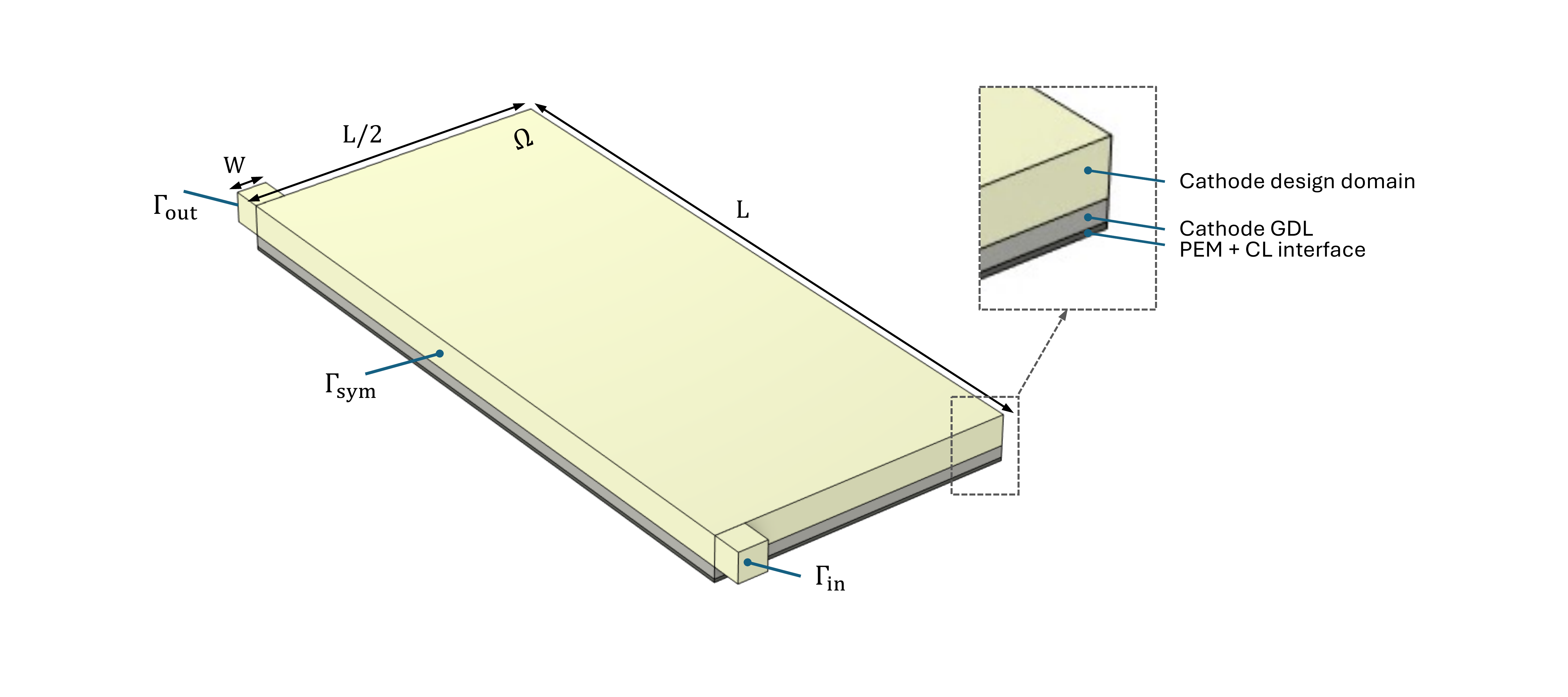}
    \caption{PEMFC cathode structural layout and optimization domain.}
    \label{fig:2}
\end{figure*}

The cathodic side of the PEMFC often limits the cell performance. This is largely because of the mass transport inefficiencies and the inherent slow kinetics of the oxygen reduction reaction (ORR). In this study, we focus on optimizing the cathode GC layout within a half-cell model depicted in Figure~\ref{fig:2}. The model consists of the cathode GDL, catalyst interface, and membrane. Humidified air enters the GCs, diffuses through the GDL with minor convective effects, and ultimately reaches the catalyst surface, where ORR occurs. The main objective is to redesign the GC geometry to enhance three-dimensional reactant transport processes, thereby improving reaction kinetics and overall fuel cell performance. Table~\ref{tab:1} lists the geometric features of the PEMFC model.

\begin{table*}[h!]
\centering
\caption{Characteristic dimensions of the PEMFC model, partly adapted from \cite{behrou2019topology}.}
\small
\setlength{\tabcolsep}{10pt}
{\begin{tabular}{l l l l}\\
\hline
{Parameter} & {Value} & {Unit} & {Description} \\
\hline
$L$ & $22$ & $\mathrm{mm}$ & Membrane length \\
$W_{in}$ & $1$ & $\mathrm{mm}$ & Inlet channel width \\
$W_{out}$ & $1$ & $\mathrm{mm}$ & Outlet channel width \\
$H_{bp}$ & $1$ & $\mathrm{mm}$ & Bipolar plate depth \\
$H_{gdl}$ & $0.38$ & $\mathrm{mm}$ & GDL thickness \\
$H_{mem}$ & $0.1$ & $\mathrm{mm}$ & Membrane depth \\
$H_{cl}$ & $0.05$ & $\mathrm{mm}$ & CL thickness \\
\hline
\end{tabular}}
\label{tab:1}
\end{table*}

\subsection{Governing equations}

The PEMFC is modeled under (i) steady-state conditions, (ii) gas flow obeys the ideal gas law and is treated as incompressible, (iii) the GDL is assumed as a homogeneous and isotropic porous material, and (iv) liquid water saturation and transport are neglected, consistent with well-humidified conditions. Based on these hypotheses, the continuity equation for the GCs and the porous medium is written as,

\begin{equation}
\rho \frac{\partial u_i}{\partial x_i}=0,
\end{equation}

\noindent where $\rho$ is the gas density and $u_i$ stands for velocity in the $x_i$ spatial direction. The momentum balance is given by,

\begin{equation}
\frac{1}{\varepsilon^2} \rho u_j \frac{\partial u_i}{\partial x_j} = -\frac{\partial p}{\partial x_i} + \frac{1}{\varepsilon} \frac{\partial}{\partial x_j} \left(\mu \left( \frac{\partial u_i}{\partial x_j} + \frac{\partial u_j}{\partial x_i} \right) \right) + F_{b_i},
\end{equation}

\noindent where $p$, $\mu$, and $\varepsilon$ represent the pressure field, gas dynamic viscosity, and porosity of the medium, respectively, with $\varepsilon$ set to $1$ in the GC domain. The term $F_{b_i}$, known as fluid friction force, is modeled as a Brinkman sink term \cite{borrvall2003topology},

\begin{equation}\label{eq:Fb}
F_{b_i} = -\alpha_b u_i.
\end{equation}

\noindent Here, $\alpha_b$ is the Brinkman constant, defined as $\alpha_b = \mu / \kappa$ with the porous material permeability $\kappa$. The species transport inside the cell is described by,

\begin{equation}\label{eq:adr}
\frac{\partial j_{k_i}}{\partial x_i} + \rho u_i \frac{\partial \omega_k}{\partial x_i} = R_{k},
\end{equation}

\noindent in which $\omega_k$ represent the species mass fraction ($k \in \{\mathrm{O}_2, \mathrm{N}_2, \mathrm{H}_2\mathrm{O}\}$). The corresponding diffusive mass flux in the $i$-th direction, $j{k_i}$, is approximated as follows,

\begin{equation}
j_{k_i}= \rho \,f_D \, D_k^m \frac{\partial \omega_k}{\partial x_i} + \rho \,f_D \, \omega_k D_k^m \frac{1}{M^m} \frac{\partial M^m}{\partial x_i},
\end{equation}

\noindent where $f_D$ is the diffusivity factor, estimated using the Bruggeman approximation as $f_D=\varepsilon^{1.5}$ for the porous GDL. Assuming isobaric and isothermal conditions, $D_k^m$ is derived based on Maxwell–Stefan equations \cite{kee2005chemically},

\begin{equation}
D_k^m = \frac{1 - \omega_k}{\sum_{\substack{j \ne k}}^{n_s} \frac{\omega_j}{D_{kj}}},
\end{equation}

\noindent with the number of species $n_s$ and the diffusivity factor $D_{kj}$, computed as follows \cite{cussler2009diffusion, cetinbas2014three, um2000computational},

\begin{equation}
D_{kj} = D_{kj}^{\text{ref}} \left( \frac{p_{\text{ref}}}{p} \right) \left( \frac{T}{T_{\text{ref}}} \right)^{1.75}.
\end{equation}

The average molar mass $M^m$ is given by \cite{cetinbas2014three}, 

\begin{equation}
\frac{1}{M^m} = \sum_{k=1}^{n_s} \frac{\omega_k}{M_k}.
\end{equation}

The reaction term $R_k$ in Eq.~\ref{eq:adr} represents the mass generation rate of $O_2$ and $H_2O$ in the electrochemical reactions \cite{um2000computational},

\begin{equation}
R_i = \frac{M_i \, \nu_i \, a_v \, i_{loc}}{n_e F}, 
\end{equation}

\noindent where $a_v$ is the electrochemically active surface area, $n_e$ stands for the electron transfer number in the ORR, $\nu_i$ is the stoichiometric coefficient of the $i$-th species involved in the reaction, and $F$ refers to the Faraday constant. The reaction rate is correlated with the local current density $i_{\text{loc}}$ that follows the Butler-Volmer equation \cite{ganser2019extended, cetinbas2014three},

\begin{equation}
i_{\text{loc}} = i_0 \left( \exp\left( \frac{\alpha_a F \eta_c}{R_g \, T} \right) - \exp\left( \frac{-\alpha_c F \eta_c}{R_g \, T} \right) \right),
\end{equation}

\noindent in which $\alpha_a$ and $\alpha_c$ are the transfer parameters of the anode and cathode, respectively, with $\alpha_a=n_e-\alpha_c$. The term $i_0$ represents the volumetric exchange current density, and $R_g$ refers to the universal gas constant. The cathodic overpotential $\eta_c$ is given by \cite{ganser2019extended, um2000computational},

\begin{equation}
\eta_c = V_{eq} - V_{cell} - R \, i_{loc},
\end{equation}


\noindent where $V_{eq}$, $V_{cell}$, and $R$ are the equilibrium voltage, the applied cell voltage, and the resistance associated with charge transport through the cell, respectively. At the $\Gamma_{in}$ boundary shown in Figure~\ref{fig:2}, a normal flow condition is prescribed such that,

\begin{equation}
u \cdot n = u_{in} \quad \quad \mathrm{on \; \; \, \Gamma_{in}},
\end{equation}

\noindent where $n$ is the outward unit normal vector. The interior walls are subjected to a no-slip condition, and the axis of symmetry is constrained by a non-penetration condition.

\begin{equation}
u_i = 0 \quad \quad \mathrm{on \; \; \, \Gamma_{tot}/(\Gamma_{in} \, \cup \, \Gamma_{out} \, \cup \, \Gamma_{sym})},
\end{equation}

\begin{equation}
u_i n_i = 0 \quad \quad \mathrm{on \; \; \, \Gamma_{sym}}.
\end{equation}

The outlet boundary is subjected to a zero-stress condition,

\begin{equation}
\left[ -p \, \delta_{ij} + \eta \left( \frac{\partial u_i}{\partial x_j} + \frac{\partial u_j}{\partial x_i} \right) \right] n_j = 0
\quad \text{on } \Gamma_{\text{out}}
\end{equation}

\noindent where $\delta_{ij}$ denotes the Kronecker delta. The physical constants and modeling parameters are listed in Table~\ref{tab:3}.

\begin{table*}[h!]
\centering
\caption{Relevant model parameters and physical constants, adapted from \cite{haghayegh2017modeling}.}
\small
\setlength{\tabcolsep}{10pt}
{\begin{tabular}{l l l l}\\
\hline
{Parameter} & {Value} & {Unit} & {Description} \\
\hline
$T$ & 343.15 & $\mathrm{K}$ & Cell operating temperature \\
$T_{ref}$ & 298.15 & $\mathrm{K}$ & Reference temperature \\
$P_{ref}$ & 1 & $\mathrm{atm}$ & Reference pressure \\
$R_g$ & 8.3144 & $\mathrm{J/(mol \cdot K)}$ & Universal gas constant \\
$F$ & 96485.3 & $\mathrm{C/mol}$ & Faraday constant \\
$V_{in}$ & 0.2 & $\mathrm{m/s}$ & Inlet flow velocity \\
$\mu$ & $2.07 \times 10^{-5}$ & $\mathrm{Pa \cdot s}$ & Gas dynamic viscosity \\
$\rho$ & 1.142 & $\mathrm{kg/m^3}$ & Gas density \\
$\varepsilon_{gdl}$ & 0.8 & - & GDL Porosity \\
$\kappa_{gdl}$ & $5 \times 10^{-11}$ & $\mathrm{m^2}$ & GDL Permeability \\
$D_{O_2,\, N_2}^{ref}$ & $2.07 \times 10^{-5}$ & $\mathrm{m^2/s}$ & Oxygen diffusivity in nitrogen \\
$D_{O_2,\, H_2O}^{ref}$ & $2.64 \times 10^{-5}$ & $\mathrm{m^2/s}$ & Oxygen diffusivity in water \\
$D_{N_2,\, H_2O}^{ref}$ & $2.64 \times 10^{-5}$ & $\mathrm{m^2/s}$ & Nitrogen diffusivity in water \\
$V_{cell}$ & 0.7 & $\mathrm{V}$ & Prescribed cell voltage \\
$V_{oc}$ & 1.22 & $\mathrm{V}$ & Open-circuit voltage \\
$i_0$ & 0.17 & $\mathrm{A/m^2}$ & Volumetric exchange current density \\
$\alpha_c$ & 0.5 & - & Cathodic transfer coefficient \\
$n_e$ & 4 & - & Number of participating electrons \\
$C_{O_2}^{\text{ref}}$ & 30 & $\mathrm{mol/m^3}$ & Reference oxygen concentration \\
$\omega_{O_2}^{in}$ & 0.228 & - & Intake oxygen mass fraction at inlet \\
$\omega_{H_2O}^{in}$ & 0.023 & - & Intake water vapor mass fraction at inlet \\
$\omega_{N_2}^{in}$ & 0.749 & - & Intake nitrogen mass fraction at inlet \\
$M_{O_2}$ & 0.032 & $\mathrm{kg/mol}$ & Oxygen molar mass \\
$M_{N_2}$ & 0.028 & $\mathrm{kg/mol}$ & Nitrogen molar mass \\
$M_{H_2O}$ & 0.018 & $\mathrm{kg/mol}$ & Water vapor molar mass \\
$n_s$ & 3 & - & Number of species ($O_2$, $N_2$, $H_2O$) \\
$\nu_{N_2}$ & $0$ & - & Stoichiometric coefficient for $N_2$ \\
$\nu_{O_2}$ & $-1$ & - & Stoichiometric coefficient for $O_2$ \\
$\nu_{H_2O}$ & $2$ & - & Stoichiometric coefficient for $H_2O$ \\

\hline
\end{tabular}}
\label{tab:3}
\end{table*}

\section{Topology optimization}\label{sec:3}

\subsection{Material model}
A density-based topology optimization approach is adopted, wherein the originally discrete design variable, $\gamma \in \{0, 1\}$, is relaxed into a continuous one, $\gamma \in [0, 1]$, and makes gradient-based optimization possible and efficient. The material indicator $\gamma$ assumes intermediate values, where $\gamma = 1$ corresponds to the gas channel region and $\gamma = 0$ represents the rib region. The interpolation laws are formulated so that the non-physical intermediate values of $\gamma$ are penalized to drive a physically meaningful, near-binary layout. Within this framework, the design-dependent physical properties are interpolated across the domain $\Omega$ as functions of $\gamma$. In Equation~\ref{eq:Fb}, the body force term that accounts for the fluid friction force, the Brinkman constant $\alpha_b(\gamma)$ is interpolated using the rational approximation of material properties (RAMP) scheme, a robust and widely used interpolation method in density-based topology optimization \cite{alexandersen2014topology, stolpe2001alternative}, and is given by,

\begin{equation}
\alpha_b(\gamma) = \frac{\mu}{\kappa_{\text{c}}} + \left( \frac{\mu}{\kappa_{\text{r}}} - \frac{\mu}{\kappa_{\text{c}}} \right) \frac{1 - \gamma}{1 + q_{\alpha} \gamma},
\end{equation}


\noindent here $q_{\alpha}$ is the convexity factor that controls how sharp the interpolation transition is between the solid ($\gamma=0$) and fluid ($\gamma=1$) areas. The parameter $\kappa_{\text{c}}$ is the channel permeability, and $\kappa_{\text{r}}$ corresponds to the rib permeability. Given that the channel is fully open, its permeability is infinite, which means the viscous resistance term $\mu/\kappa_{\text{c}}$ approaches zero. The diffusivity correction factor is interpolated based on solid isotropic material with penalization (SIMP) style, as described in \cite{rietz2001sufficiency},

\begin{equation}
f_D(\gamma) = f_{D_r}+\left(f_{D_c}-f_{D_r} \right) \, \gamma^{q_f},
\end{equation}

\noindent where $f_{D_r}$ corresponds to the rib diffusivity factor and $f_{D_c}$ is that of the channel. The diffusivity of the rib is assumed to be zero. The diffusivity factor of the porous GDL is given by the Bruggeman correlation $f_D = \varepsilon^{1.5}$, as previously mentioned. The parameter $q_f$ refers to the SIMP penalization exponent.

\subsection{Reduced-order model}
The three-dimensional multilayered structure, depicted in Figure~\ref{fig:2}, is reduced into a representative plane for two-dimensional topology optimization. This reduced model can account for the in-plane reactant convection and diffusion, but it is important to retain the out-of-plane species transport into the adjacent GDL. To approximate this effect, the GDL is integrated into the bipolar plate through a depth-averaging method following \cite{he2019reduced}. In this simplified model, contact resistance between layers is overlooked. Equivalent transport properties of the representative plane, namely, permeability and the diffusivity correction factor, are computed for both channel and rib regions using a depth-weighted mean. Porosity is treated similarly. The equivalent property is given by,

\begin{equation}
\overline{(\cdot)}_c = \frac{(\cdot)_{\mathrm{gdl}} H_{\mathrm{gdl}} + (\cdot)_{\mathrm{c}} H_{\mathrm{bp}}}{H_{\mathrm{gdl}} + H_{\mathrm{bp}}},
\end{equation}

\begin{equation}
\overline{(\cdot)}_r = \frac{(\cdot)_{\mathrm{gdl}} H_{\mathrm{gdl}} + (\cdot)_{\mathrm{r}} H_{\mathrm{bp}}}{H_{\mathrm{gdl}} + H_{\mathrm{bp}}},
\end{equation}

\noindent where $\overline{(\cdot)}_c$ ($\overline{(\cdot)}_r$) correspond to the channel (rib) depth-averaged property. $H_{bp}$ and $H_{gdl}$ are the thicknesses of the GDL and bipolar plate, respectively. These approximated properties are then used in interpolation in Section~\ref{sec:3} for two-dimensional topology optimization in the representative plane. In this context, the local porosity $\varepsilon (\gamma)$ is also interpolated as follows, 

\begin{equation}
\varepsilon(\gamma) = \varepsilon_{r} + (\varepsilon_{c} - \varepsilon_{r}) \, \gamma,
\end{equation}

\noindent where $\varepsilon_c$ is the depth-averaged porosity of the channel, and $\varepsilon_r$ that of the rib area.

\subsection{Objectives and constraints}

Our objective is to maximize the reactant concentration over the catalyst sites while, at the same time, the most homogeneous spatial distribution of the reactant is created with low pressure drop and power dissipation along the flow path. This is especially important on the cathode side, where a higher reactant concentration accelerates the ORR and leads to increased current density generation. In parallel, a more homogeneous reactant distribution helps prevent localized reactant starvation and the formation of hotspots. Accordingly, our optimization objective is the maximization of the reactant concentration at a minimized pressure drop. The latter is typically quantified by the power dissipation function $J_f$ over $\Omega$. These objectives are combined into a weighted sum formulation,

\begin{equation}
\begin{aligned}
\max_{\boldsymbol{\gamma} \in \mathcal{A}} \quad & \omega \, \frac{\overline{C}}{\overline{C}_{\mathrm{init}}}-(1-\omega) \, \frac{J}{J_{\mathrm{init}}} \\
\text{subject to} \quad & \int_{\Omega} \gamma(\mathbf{x}) \, \mathrm{d}\mathbf{x} \leq V^{\star}\\
\quad & \mathcal{A} = \left\{ \boldsymbol{\gamma} \in \mathbb{R}^{N_{\text{dof}}} \;\middle|\; 0 \leq \gamma_n \leq 1, \; n = 1, \dots, N_{\text{dof}} \right\},
\end{aligned}
\end{equation} 

\noindent and $\mathbf{\gamma}$ is a vector of $N_{\text{dof}}$ design variables, with each $\gamma_n$ bounded between $0$ and $1$. The parameter $\omega$ denotes the weighting factor. The volume taken by the gas channels is constrained to remain below the prescribed threshold $V^{\star}$ and prevent trivial or undesirable design solutions. The power dissipation $J$ is defined as,

\begin{equation}
J = \int_{\Gamma} \left( \frac{1}{2} \mu \sum_{i,j} \frac{\partial u_i}{\partial x_j} \left( \frac{\partial u_i}{\partial x_j} + \frac{\partial u_j}{\partial x_i} \right) + \alpha_b \sum_{i} u_i^2 \right) \, \mathrm{d}\Gamma,
\end{equation}

\noindent and the mean reactant concentration is given by,

\begin{equation}
\overline{C} = \frac{\int_{\Omega} C \, \mathrm{d}\mathbf{x}}{\int_{\Omega} \mathrm{d}\mathbf{x}}.
\end{equation}

\noindent Because of the difference in scale between the mean reactant concentration and the flow resistance, both quantities are normalized. ${\overline{C}_{\mathrm{init}}}$ and ${J_{\mathrm{init}}}$ represent the initial values of the mean concentration and the power dissipation of the concentration field, respectively, measured for a domain with uniformly distributed material $\gamma = V^{\star}$.

\subsection{Filtering and regularization}
Density-based topology optimization suffers from numerical issues such as checkerboard patterns, sensitivity to mesh resolution, and the presence of non-physical, blurry intermediate materials. Such instabilities can be resolved with regularization approaches. Filtering schemes are widely used to mitigate such instabilities and promote discrete and manufacturable design solutions. The Helmholtz filter \cite{lazarov2011filters} is applied to address the checkerboard phenomenon, 

\begin{equation} 
-r_{\text{min}}^2 \nabla^2 \gamma_f + \gamma_f=\gamma.
\end{equation}

The variable $\gamma_f$ represents the filtered (smoothed) design field, and $r_{min}$ is the filter radius, which defines the minimum feature size of the structure and controls the design resolution. To maintain numerical stability, the filter radius must not be smaller than the mesh size. While filtering mitigates numerical artifacts such as checkerboard patterns and mesh dependency, it often leaves nonphysical grayscale regions, which complicates the interpretation and fabrication of the optimized layout. To reduce such blurriness, a heaviside projection method (HPM) based on the tanh function \cite{wang2011projection} is adopted as a post-filtering step,

\begin{equation}
\gamma_h = \frac{\tanh(\beta \, \theta_p) + \tanh(\beta \, (\gamma_f - \theta_p))}{\tanh(\beta \, \theta_p) + \tanh(\beta \, (1 - \theta_p))},
\end{equation}

\noindent where $\gamma_h$ and $\beta$ are the projected (binarized) design variable and the projection cutoff, respectively, and $\theta_p$ controls the projection slope. A higher value of $\beta$ results in a more pronounced distinction between solid and fluid domains.The optimization parameters outlined in Table~\ref{tab:2} were derived based on exploratory analysis specific to this problem.

\begin{table*}[h!]
\centering
\caption{List of parameters utilized in the optimization problem.}
\small
\setlength{\tabcolsep}{10pt}
{\begin{tabular}{l l l l}\\
\hline
{Parameter} & {Value} & {Unit} & {Description} \\
\hline
$q_{\alpha}$ & 0.04 & - & Convexity factor for interpolation \\
$q_{f}$ & 3 & - & SIMP penalization exponent \\
$r_{min}$ & 0.25 & $\mathrm{mm}$ & Filter radius \\
$\beta$ & 2 & - & Slope of Heaviside projection function \\
$\theta_p$ & 0.5 & - & Projection threshold \\
$V^{*}$ & 0.5 & - & Maximum allowable volume fraction \\
$C_{i}$ & 1000 & - & Constraint penalty coefficient \\
$t_{opt}$ & $1 \times 10^{-5}$ & - & Optimality convergence tolerance \\
$N_{ev}$ & 80 & - & Maximum number of model evaluations \\
$N_{in}$ & 10 & - & Maximum number of inner iterations \\
\hline
\end{tabular}}
\label{tab:2}
\end{table*}

\subsection{Optimization setup}
\begin{figure*}[h!]
    \centering
    \includegraphics[width=13cm]{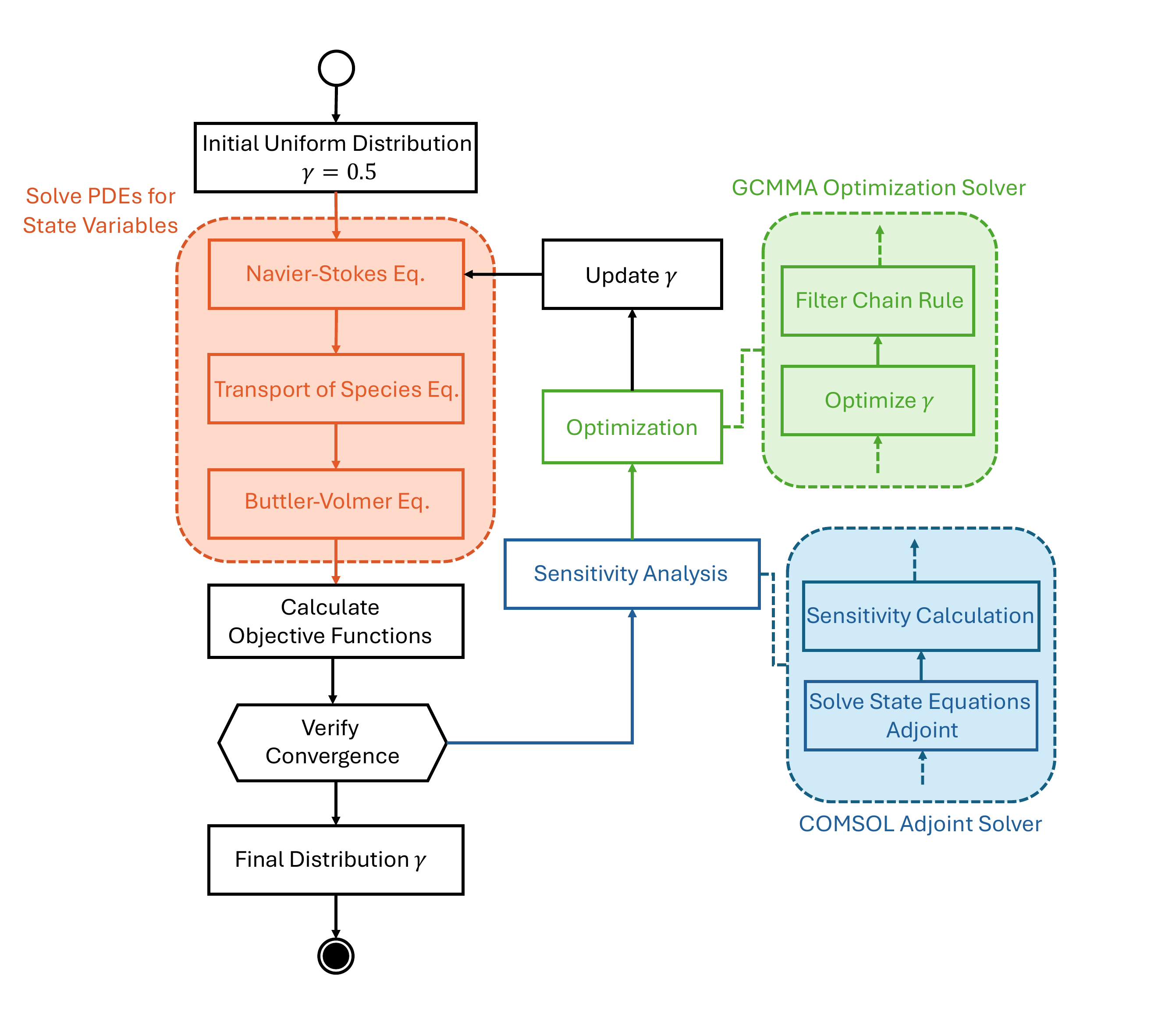}
    \caption{Flow diagram of the optimization process consisting of primal, adjoint, and optimization solvers.}
    \label{fig:3}
\end{figure*}

COMSOL Multiphysics\textsuperscript{\textregistered} 6.2 is used to solve the optimization problem. The computational domain is discretized with structured linear hexahedral elements in 3D and quadrilateral elements in 2D. Characteristic mesh sizes of $\Delta h=0.16$ mm (3D) and $\Delta h=0.07$ (2D) are selected based on a mesh convergence study on an optimization case, in which the dissipation energy along the flow path is evaluated for five different grid resolutions. The grid independence test results in Table~\ref{tab:4} show that further mesh refinement has a negligible effect on the solution and confirm mesh-independent behavior. The characteristic mesh size is estimated as follows,

\begin{equation}
\Delta h = \left(\frac{V}{N_{\text{ele}}}\right)^{\frac{1}{d}},
\end{equation}

\noindent where $V$ is the meshed volume, $N_{ele}$ is the number of mesh elements, and $d$ represents the spatial dimensionality of the problem. Finite element analysis (FEA) is performed to solve the governing equations for the primal state variables of the pressure, velocity, and species mass fractions. These variables are estimated by linear interpolation functions. The system of nonlinear equations is solved by Newton's method. At each Newton iteration, the linear system is handled using the multifrontal massively parallel sparse direct solver (MUMPS) \cite{amestoy2001fully}. Convergence is assumed when the residual norm reaches below $10^{-3}$. Sensitivities are computed using automatic differentiation in COMSOL. The adjoint method is used to analyze the objective and constraint sensitivities to the design variable $\gamma$. These sensitivities are then passed to the globally convergent method of moving asymptotes (GCMMA) optimizer \cite{svanberg2002class} to iteratively update the material distribution. The optimization terminates when either the variation in the design variables is less than $10^{-6}$ or once $N_{eval}$ evaluations are reached. The main steps of the optimization process are outlined in Figure~\ref{fig:3}.

\begin{table*}[h!]
\centering
\caption{The grid independence test on an optimization case.}
\small
\setlength{\tabcolsep}{10pt}
{\begin{tabular}{l l l l l l}\\
\hline
\multicolumn{3}{c}{2D} & \multicolumn{3}{c}{3D} \\
\hline
{$\Delta{h}$ [mm]} & $J_f$ [$\mathrm{W/m^3}$] & Relative error [\%] & {$\Delta{h}$ [mm]} & $J_f$ [$\mathrm{W/m^3}$] & Relative error [\%] \\
\hline
0.19 & 17.515 & 30.8 & 0.44 & 4.256 & 66.5\\
0.16 & 25.338 & 27.1 & 0.27 & 12.699 & 24.7 \\
0.13 & 34.751 & 8.9 & 0.19 & 16.877 & 8.6 \\
0.10 & 38.115 & 2.2 & 0.16 & 18.456 & 3.0 \\
0.07 & 38.975 & - & 0.13 & 19.027 & - \\
\hline
\end{tabular}}
\label{tab:4}
\end{table*}

\section{Results and discussion}
\subsection{Model validation}
\begin{figure*}[h!]
    \centering
    \includegraphics[width=13cm]{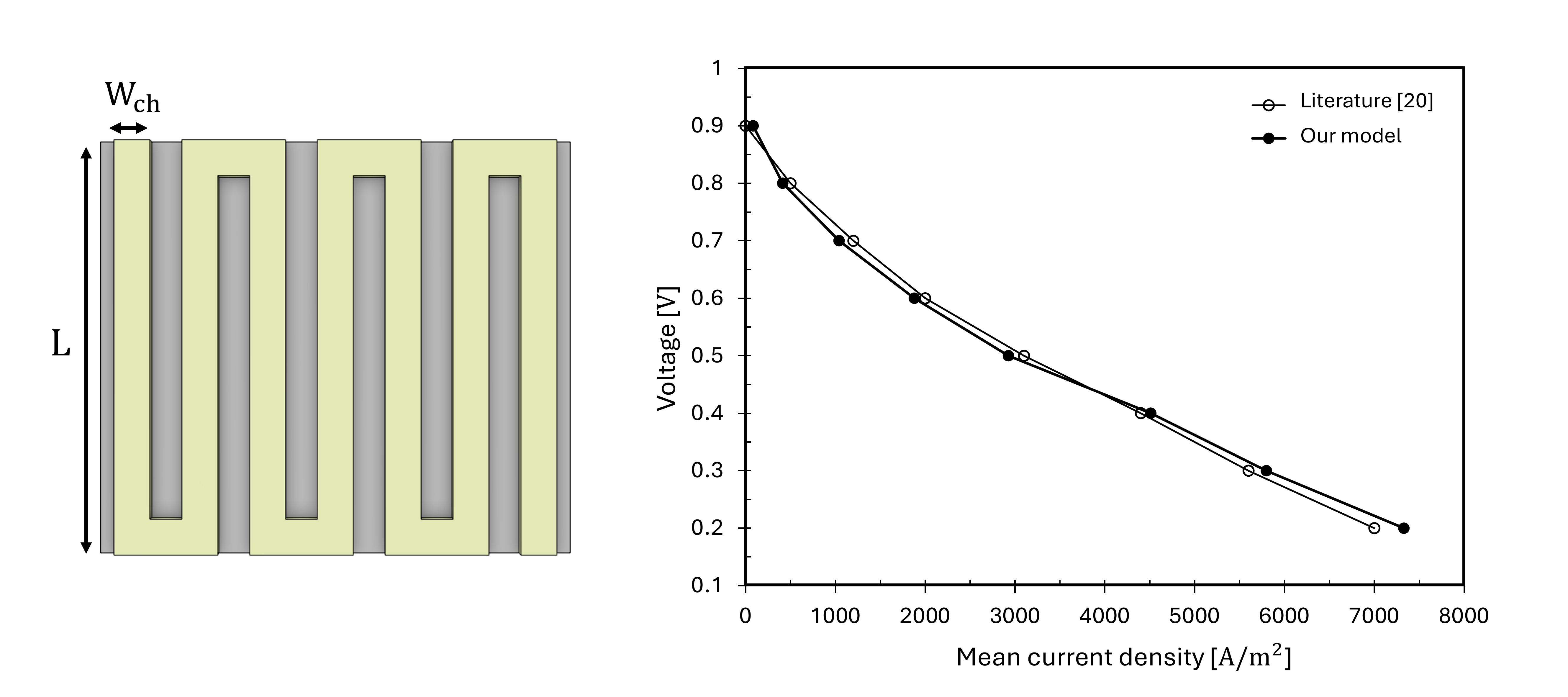}
    \caption{Polarization curves predicted by our PEMFC model, verified against the numerical results reported in \cite{haghayegh2017modeling}.}
    \label{fig:10}
\end{figure*}

The PEMFC model is validated against the study of Haghayegh et al. \cite{haghayegh2017modeling}, on a single serpentine channel configuration shown on the right side of Figure~\ref{fig:10}. The geometric specifications of the model are presented in Table~\ref {tab:9}, and the physical and material properties are adopted from Table~\ref{tab:3}, consistent with the reference study. The system is fed by an oxygen–water mixture at a flow rate of $2 \times 10^{-6}$ $\mathrm{m^3/s}$, consistent with the conditions reported in \cite{haghayegh2017modeling}. The computational domain is discretized using structured hexahedral elements. In the reference study \cite{haghayegh2017modeling}, the anode electrochemistry is described by the Butler–Volmer equation, while the cathode reaction is modeled with a simplified Tafel equation. Since anode electrochemistry is not explicitly included in our PEMFC model, the exchange current density is calibrated so that the model aligns with the reference results in \cite{haghayegh2017modeling}. Following the curve-fitting procedure in \cite{behrou2019topology}, this calibration gives $i_0 = 0.17$ $\mathrm{A/m^2}$. The resulting polarization curve from our 3D model is verified with the reference results in Figure~\ref{fig:10}. The comparison shows good agreement with a mean relative deviation of only $3.2\%$. In general, reasonable agreement is observed between our model predictions and the reference results of \cite{haghayegh2017modeling}, which demonstrates that the half-cell PEMFC model reliably captures the electrochemical behavior of the cell with sufficient precision.

\begin{table*}[h!]
\centering
\caption{Geometric parameters of the reference PEMFC model, adapted from \cite{haghayegh2017modeling}.}
\small
\setlength{\tabcolsep}{10pt}
{\begin{tabular}{l l l l}\\
\hline
{Parameter} & {Value} & {Unit} & {Description} \\
\hline
$L$ & 20.82 & $\mathrm{mm}$ & Channel length \\
$H_{ch}$ & 1.8 & $\mathrm{mm}$ & Channel thickness \\
$W_{ch}$ & 1.8 & $\mathrm{mm}$ & Channel width \\
$W_{r}$ & 1.6 & $\mathrm{mm}$ & Rib width \\
$H_{gdl}$ & 0.2 & $\mathrm{mm}$ & GDL thickness \\
$H_{cl}$ & 0.05 & $\mathrm{mm}$ & CL thickness \\
$H_{mem}$ & 0.21 & $\mathrm{mm}$ & Membrane thickness \\
$N_{ch}$ & 7 & - & Number of channels \\
$V_{in}$ & 0.62 & $\mathrm{m/s}$ & Cathode inlet flow rate \\
$P$ & 3 & $\mathrm{atm}$ & Operation pressure \\
$P_{ref}$ & 1 & $\mathrm{atm}$ & Reference pressure \\
$T$ & 343.15 & $\mathrm{K}$ & Cell temperature \\
\hline
\end{tabular}}
\label{tab:9}
\end{table*}

\subsection{Three-dimensional optimized GCs}\label{sec:4.1}
\begin{figure*}[h!]
    \centering
    \includegraphics[width=18cm]{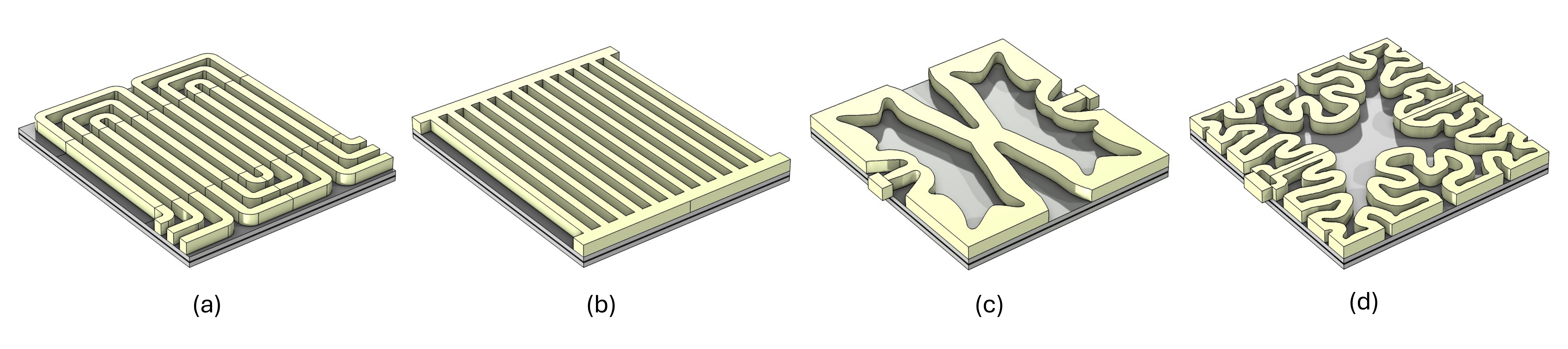}
    \caption{Three-dimensional GC structures: (a) conventional serpentine layout, (b) conventional parallel layout, (c) topology-optimized design derived in three dimensions, and (d) three-dimensional extrusion of the reduced-order, two-dimensional topology-optimized design.}
    \label{fig:0}
\end{figure*}

\begin{figure*}[h!]
    \centering
    \includegraphics[width=15cm]{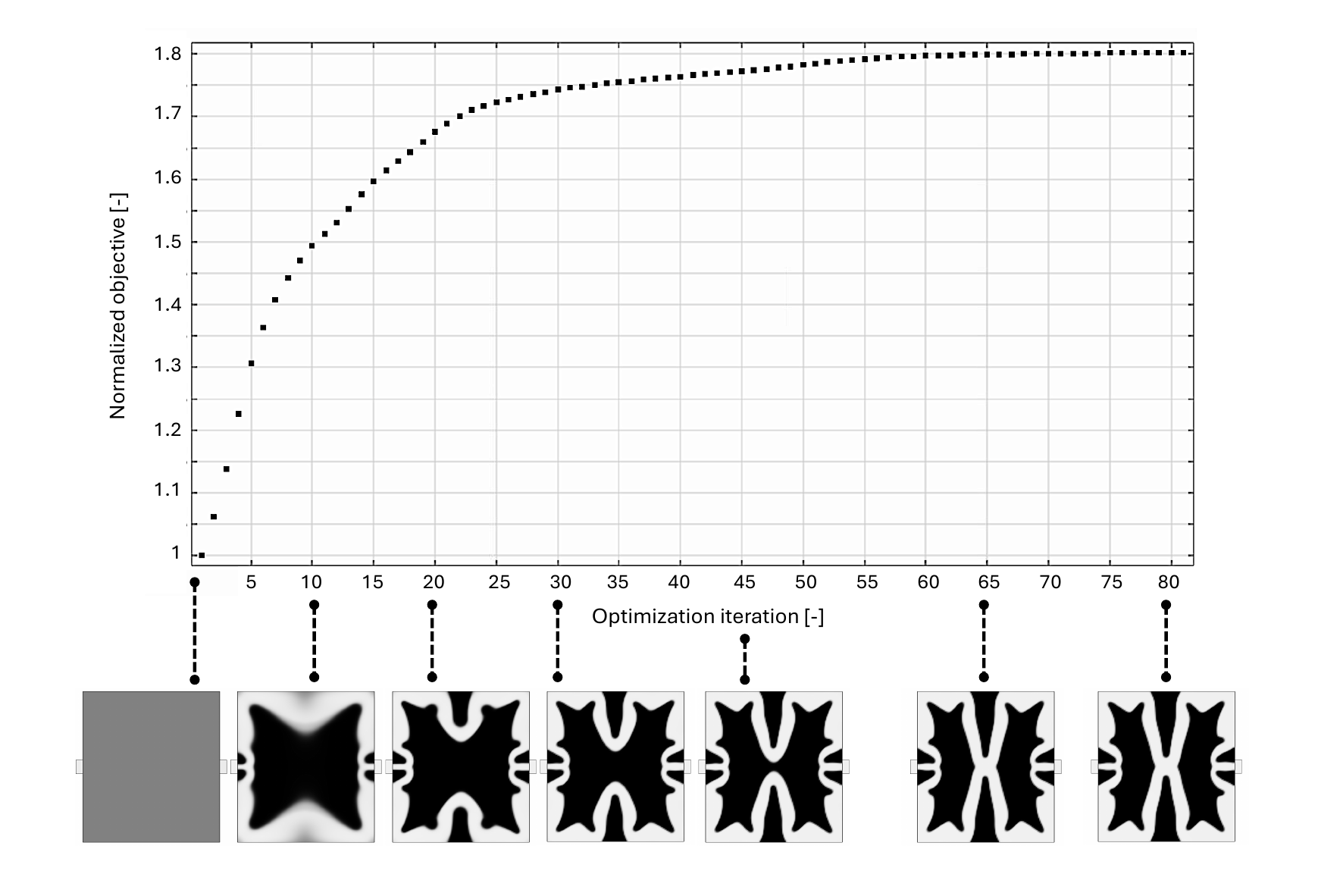}
    \caption{Objective history along with representations of the projected material distribution.}
    \label{fig:5}
\end{figure*}

The biobjective optimization problem is solved in three dimensions to optimize the cathode GC layout within the half-cell PEMFC model shown in Figure~\ref{fig:2}. This section presents the optimization solution for a reference case with an objective weight factor $\omega$ of $0.8$ and an input flow rate $u_{in}$ of $0.1 \, \mathrm{m/s}$ at a potential of $0.7 \,\mathrm{V}$ applied to the cell. The optimized design is then compared to typical three-dimensional benchmark serpentine and parallel layouts in Figure~\ref{fig:0}(a) and (b). The benchmarks use the same active area and domain dimensions as the optimization domain. The width and number of channels are selected so that the volume constraint $V^*$ is satisfied to ensure a meaningful comparison. A fixed volumetric flow rate of $0.3$ $\mathrm{cm^3/s}$ is assumed at the inlet, with oxygen, water vapor, and nitrogen mass fractions of $0.228$, $0.023$, and $0.749$, respectively. A stress-free condition is applied at the outlet boundaries. All other relevant modeling and optimization parameters, listed in Table~\ref{tab:2} and Table~\ref{tab:3}, are shared between the topology-optimized and benchmark designs. Figure~\ref{fig:5} shows how the objective evolves and converges, along with the projected material layout. As optimization progresses, the objective initially rises rapidly and then gradually levels off. The optimized structure develops from an initially even material indicator field, $\gamma=0.5$, into a configuration where material builds up near the walls and forms two main channels that connect the inlet to the outlet. As optimization continues, a connection between these two main channels starts to appear in the middle of the domain and forms an X-shaped pattern. Notably, there are very few intermediate blurred regions, and the design converges to a nearly binary pattern, which indicates effective regularization.

\begin{figure*}[h!]
    \centering
    \includegraphics[width=17cm]{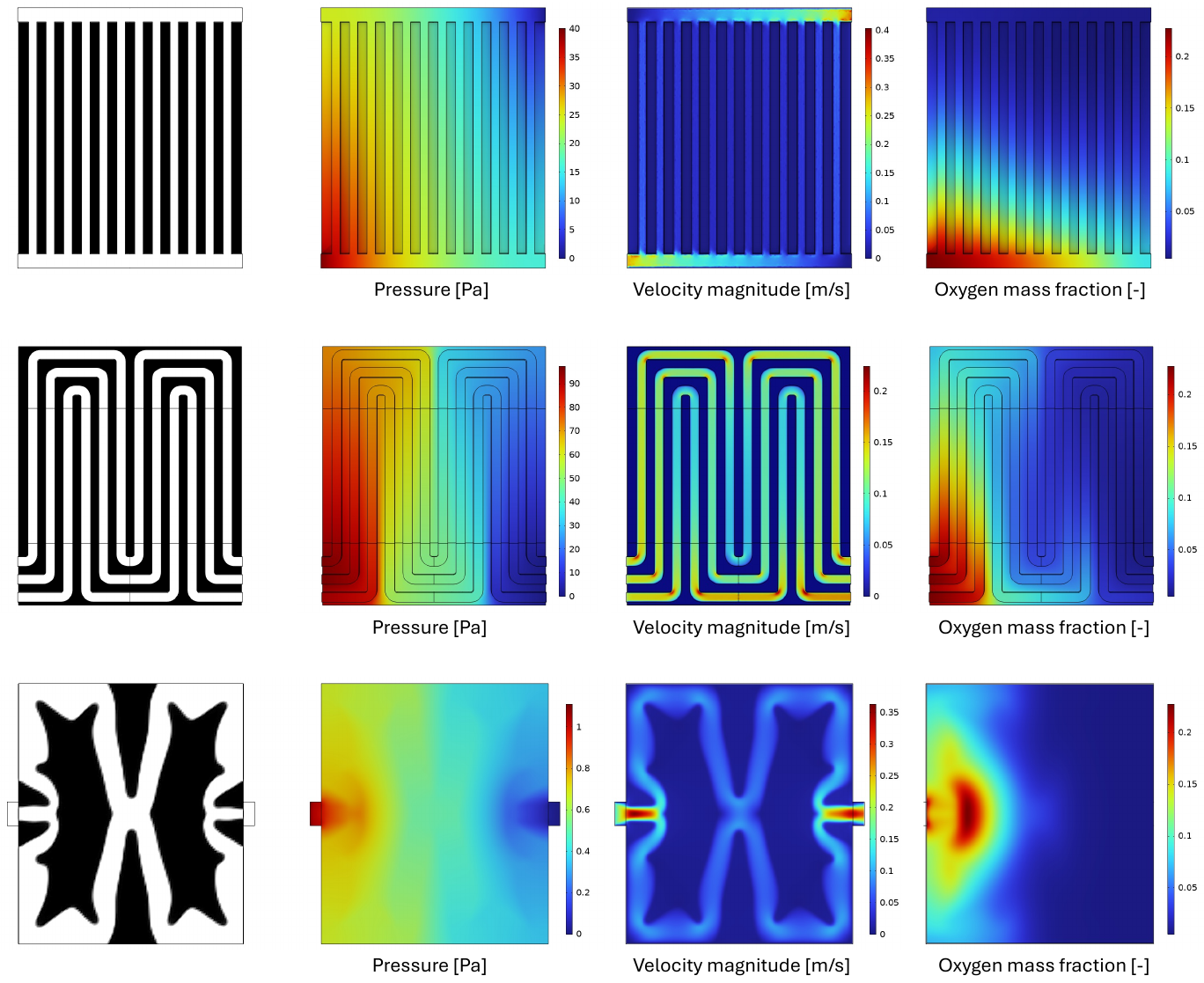}
    \caption{Contours of pressure, velocity, and oxygen concentration in the three-dimensional benchmark parallel and serpentine flow channel layouts, and the topology-optimized layout. Results correspond to an inlet volumetric flow rate of $0.3$ $\mathrm{cm^3/s}$, with the optimized topology obtained for an objective weight factor $\omega=0.8$.}
    \label{fig:4}
\end{figure*}

Figure~\ref{fig:4} presents the final optimized topology along with its pressure and velocity fields. Compared to conventional designs, the optimized layout exhibits a much lower pressure drop of only $1.1$ Pa and a reduced power dissipation of $17.95$ $\mathrm{W/m^3}$ thanks to its smoothly curved channels. However, the design lacks sub-branches and short flow paths that could enhance reactant distribution homogeneity. This limitation likely arises because 3D topology optimization is susceptible to local minima and become trapped in suboptimal design solutions or, in some cases, fail to converge. The serpentine channels span the design domain and allow a more homogenized reactant flow distribution, which results in a mean current density $i_{loc}$ of $1815.3$ $\mathrm{A/m^2}$ with a standard deviation of $1249.3$ $\mathrm{A/m^2}$ over the current collector. However, this leads to a notable pressure drop of $95.9$ Pa between the inlet and outlet. In contrast, the topology-optimized design significantly reduces the pressure drop compared to both the serpentine and parallel designs. This improvement eliminates the need for external blowers for reactant recirculation, which would otherwise add extra volume and weight to the system. Remarkably, the optimized GCs maintain a mean current density of $1728.0$ $\mathrm{A/m^2}$, which is comparable to that of serpentine design. It also outperforms the parallel design by $30.6$\% in mean current density. A summary of these comparisons is provided in Table~\ref{tab:5}. Inspection of the velocity contours shows that reactant flow is slow within the optimized gas pathways, similar to serpentine and parallel channels. This low flow speed leads to rapid oxygen consumption and depletion near the inlet (see oxygen mass fraction in Figure~\ref{fig:4}). In parallel design, uneven fluid resistance between channels also causes localized regions of severe oxygen depletion. 

\begin{table*}[h!]
\centering
\caption{Comparison of three-dimensional cathode gas channel layouts in Figure\ref{fig:0} at identical inlet volumetric flow rate.}
\begin{tabular}{lcccc}\\
\hline
Performance metric & Topology-optimized & Parallel & Serpentine & Topology-optimized (2D extruded to 3D) \\
\hline
Inlet volumetric flow rate [$\mathrm{cm^3/s}$] & $0.3$ & $0.3$ & $0.3$ & $0.3$ \\
Pressure drop [Pa]  & 1.1 & 39.9 & 95.9 & 51.1\\
Mean current density [$\mathrm{A/m^2}$] & 1728.0 & 1323.0 & 1815.3 & 2196.2\\
Current density standard deviation [$\mathrm{A/m^2}$] & 2014.7 & 1032.5 & 1249.3 & 1187.5\\
\hline
\end{tabular}
\label{tab:5}
\end{table*}

\subsection{Two-dimensional optimized GCs}
\begin{figure*}[h!]
    \centering
    \includegraphics[width=13cm]{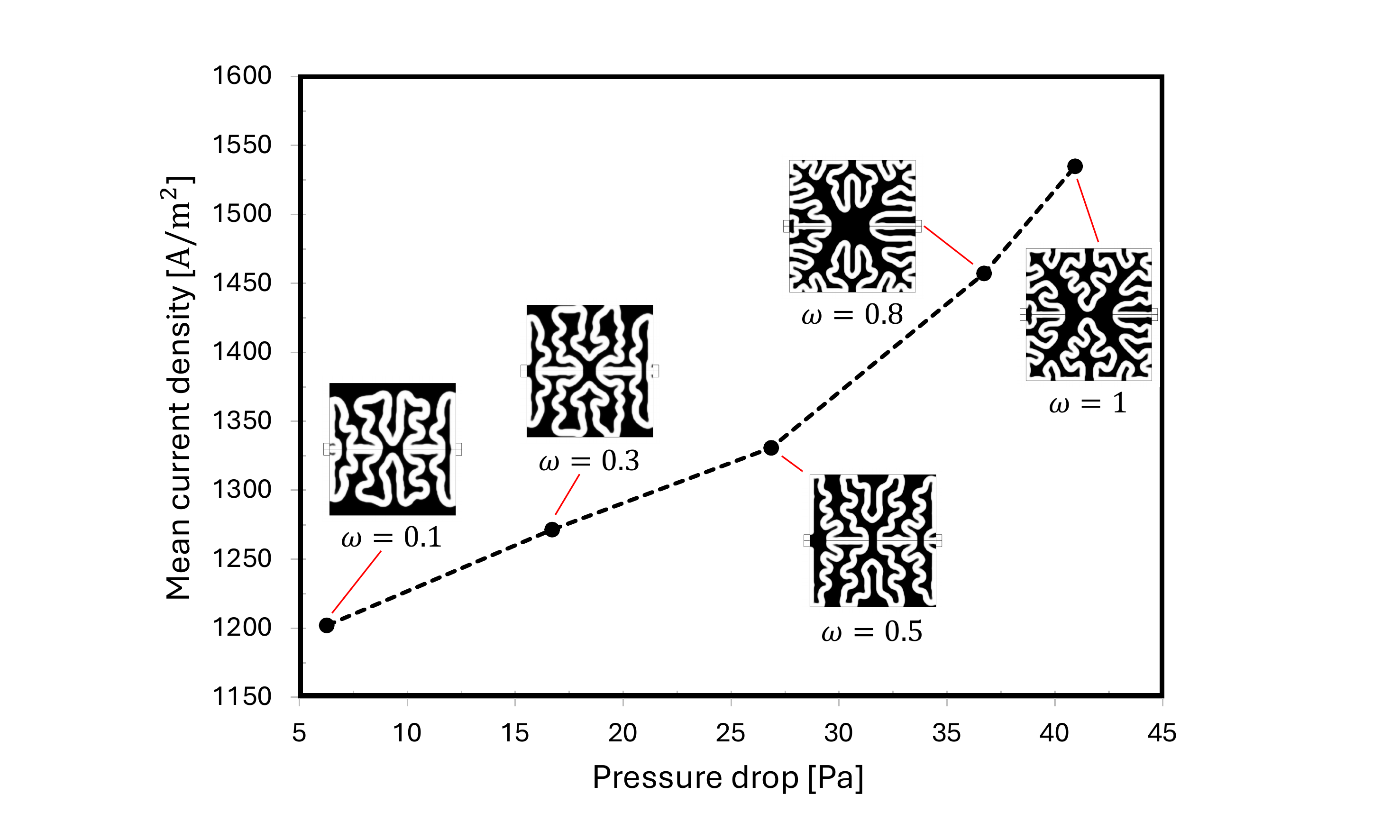}
    \caption{Graphical illustration of pressure drop and mean current density trends in optimized structures as a function of the objective weighting factor.}
    \label{fig:9}
\end{figure*}

This section discusses the optimized GC layout obtained from solving the reduced two-dimensional optimization model. The design trend by the choice of the penalty weight $\omega$ in the bi-objective formulation, at an inlet velocity of $0.1$ $\mathrm{m/s}$, is analyzed, and how it affects the physical properties is analyzed. Negligible differences are observed in the topologies shown in Figure~\ref{fig:9} for different choices of this factor. The lowest analyzed weight, $\omega=0$, gives a single thick straight path that connects the inlet to the outlet. Given that $\omega=0$ places emphasis on energy dissipation minimization, this straight channel favors the lowest pressure drop, less than $1$ Pa. The first topology in Figure~\ref{fig:9} shows the optimized flow field with $\omega=0.1$, in which tortuous pathways are formed that do not fully span the upper and lower edges of the domain. As the penalty weight further increases, the optimized topology evolves into thinner, more tortuous paths that tend to cover almost the entire design domain, in particular along the boundaries. This suggests that reactant flow can be delivered across the entire domain through the convective effect. This has a negative eﬀect on the pressure drop and energy dissipation (see Figure~\ref{fig:9}). There exists a trade-off between power dissipation and concentration homogeneity: lowering flow resistance and pressure drop leads to greater variability in reactant distribution. Compared to three-dimensional optimization, which is prone to local minima, limited domain coverage, and reactant heterogeneity, the reduced-order two-dimensional model generates layouts that span the domain more effectively. The study shows that the emphasis on dissipation minimization (lower $\omega$) results in simpler channel layouts with lower pressure drop, whereas higher values of $\omega$ prioritize reactant concentration and produce intricate geometries.


\begin{figure*}[t!]
    \centering
    \includegraphics[width=16cm]{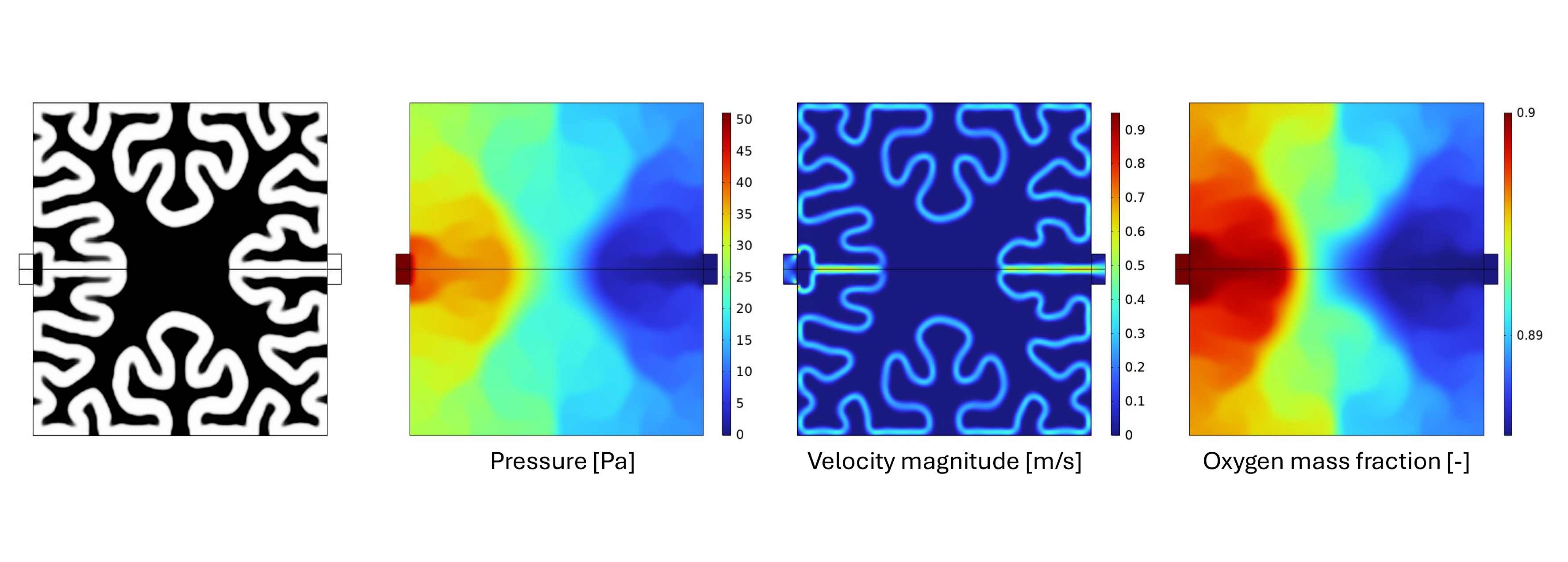}
    \caption{The two-dimensional optimized layout with the corresponding pressure, velocity, and oxygen concentration distributions, evaluated at an inlet velocity of $0.15$ $\mathrm{m/s}$ and objective weight factor $\omega =0.8$.}
    \label{fig:7}
\end{figure*}

A two-dimensional topology optimized GC, obtained from the reduced order model, is then stretched into three dimensions in the through-plane direction to conduct simulations within the half-cell model and evaluate against both the original 3D topology optimized structure and the conventional parallel and serpentine designs in Figure~\ref{fig:0}. The analysis is carried out under the same conditions as described in Section~\ref{sec:4.1}; the solid material is maintained at $50\%$, the dimensions of the active area and the design domain remain unchanged, and a fixed volumetric flow rate of $0.15,\mathrm{m}^3/\mathrm{s}$ is prescribed at the inlet. The selected optimized structure, corresponding to $\omega=0.8$, is shown in Figure~\ref{fig:7}. The channel, GDL, and CL thicknesses are identical to those of the previous section. All subsequent simulations and analyses are performed with the same parameter set. The distributions of pressure, velocity, and oxygen mass fraction are shown in Figure~\ref{fig:7}. As evidenced by the oxygen concentration contour, the oxygen distribution is more even with the standard deviation of $0.004$, compared to $0.04$ for the conventional serpentine design. According to the pressure distribution, at the input velocity $0.15$ $\mathrm{m/s}$, a pressure drop of $51.1$ Pa is observed, which is $46.7\%$ lower than that of serpentine and $28.2\%$ higher than that of the parallel channels. To further quantitatively assess the performance of the optimized layout, the velocity fields are compared. Within optimized channels, oxygen flows with higher velocity magnitudes than in the serpentine, parallel, and original 3D topology optimized cases, as shown in Figure~\ref{fig:4}. This higher velocity enables a more homogeneous delivery of reactants to the central area and closer to the outlet. The higher flow velocities also help prevent rapid localized oxygen consumption and depletion. This uniformity ensures consistent access of oxygen to the catalyst sites, efficient catalyst utilization, and mitigates the formation of hot and cold spots, thereby enhancing the cell longevity. For a more comprehensive performance assessment, the average current density and its standard deviation over the cathode current collector are computed and listed in Table~\ref{tab:5}. Compared with the serpentine case, the optimized GCs improve the mean current density by $20.9\%$ and reduce the standard deviation by $4.9\%$. This improvement is notable, given that serpentine designs are already known for the coverage and relatively uniform local current density. Compared to the original 3D optimized topology, the proposed design yields a $27.1\%$ increase in mean current density and a substantial $41\%$ reduction in standard deviation. Furthermore, it achieves a remarkable $39.7\%$ improvement in mean current density relative to the parallel layout. All the above observations demonstrate the superiority of the reduced topology-optimized design over both the conventional and the original 3D optimized alternatives. Notably, the optimized design combines a pressure drop comparable to the parallel configuration with a mean and standard deviation of current density comparable to, or better than, those of the serpentine design. In addition to its superior performance, the reduced-order approach offers improved numerical stability and computational tractability, which allows for more design exploration than direct three-dimensional topology optimization. In the selected design, the inclusion of an energy dissipation minimization term in the objective formula caused the channels not to reach the central area to lowering the pressure drop. Adjusting the weighting of this term can further reduce the current density standard deviation, but it leads to an increased pressure drop.

\section{Conclusion}

This paper shows the feasibility and potential of topology optimization to design cathode GCs in PEMFCs. This approach allows the channel to evolve freely during the optimization process and enables the creation of unconventional, non-intuitive geometries that can outperform standard alternatives. The optimization objective is to maximize reactant concentration and minimize energy dissipation at the same time. The optimization problem is solved in three dimensions within a half-cell model. Compared to conventional serpentine and parallel configurations, the optimized layout achieves a pressure drop of $1.1$ Pa, which is substantially lower than that of both benchmarks. The optimized GC maintains a mean current density of  $1728.0$ $\mathrm{A/m^2}$, comparable to the serpentine design and $31$\% higher than the parallel design. The 3D optimized design develops thick pathways with limited domain coverage, likely because 3D topology optimization tends to get trapped in local minima and suboptimal design solutions. To address this issue and improve computational tractability, a reduced-order two-dimensional model is used, where the out-of-plane reactant transport is included through a depth-averaging scheme. Topology optimization in this framework generates a unique layout with non-trivial features that span the domain more thoroughly to ensure a homogeneous reactant distribution and mitigate the formation of oxygen-depleted regions seen in the original optimized geometry, as well as in parallel and serpentine benchmarks. The resulting design outperforms the serpentine channel with a $46.7$\% reduction in pressure drop and a $20.9$\% increase in mean current density at an inlet velocity of $0.15$ $\mathrm{m/s}$. Compared to parallel channels, it delivers a $39.7$\% improvement in mean current density, though with a $28.2\%$ increase in pressure loss. Analysis further shows that oxygen flows at higher velocity magnitudes within the optimized channels, which allows a more homogeneous delivery of reactants across the domain and avoids oxygen localized oxygen depletion. The study also highlights that placing less emphasis on minimizing dissipation yields more intricate, tortuous channel pathways that enhance current generation at a higher pressure drop. Importantly, the results indicate that higher output power can be achieved even at lower input pressures. This presents a promising approach for the development of next-generation PEMFCs with improved performance and extended lifespan. Future studies will extend the optimization framework to include multiphase flow analysis and liquid water transport. Proper water management is important as an excess liquid water can cause flooding and performance losses, while insufficient hydration leads to membrane degradation. Developing a model to capture water saturation behavior and water removal mechanisms will therefore be essential. Finally, the intricate topology-optimized geometry can be additively manufactured and experimentally validated both $\textit{in situ}$ within an operating fuel cell and $\textit{ex situ}$ under controlled laboratory conditions.

\section*{Acknowledgements}
The authors thank the Natural Sciences and Engineering Research Council of Canada (NSERC) for their financial support and acknowledge the Digital Research Alliance of Canada, which made computational resources available.

\section*{Authorship contributions}
Z. Kazemi: Conceptualization, Data curation, Formal analysis, Methodology, Validation, Visualization, Writing – original draft. K. Behdinan: Conceptualization, Funding acquisition, Project administration, Resources, Supervision, Writing – review \& editing

\section*{Declaration of generative AI and AI-assisted technologies in the writing process}
While preparing this work, the authors used ChatGPT to enhance the manuscript’s language and readability. After utilizing these tools, the authors reviewed and edited the content as necessary and take full responsibility for the publication’s content.

\bibliographystyle{unsrt} 
\bibliography{example}


\end{document}